\documentclass[twocolumn,aps,showpacs,superscriptaddress,prl]{revtex4}

\usepackage{amsmath,amssymb,amsmath}
\usepackage{graphicx}
\usepackage{dcolumn}
\usepackage{bm}

\begin{document}

\title{Formation of Ultracold Heteronuclear Dimers in Electric Fields}

\author{Rosario Gonz\'alez-F\'erez}
\email{rogonzal@ugr.es}
\affiliation{Instituto 'Carlos I' de F\'{\i}sica Te\'orica y
Computacional and Departamento de F\'{\i}sica At\'omica Molecular y
Nuclear, Universidad de Granada, E-18071 Granada, Spain}

\author{Michael Mayle}
\affiliation{Theoretische Chemie, Physikalisch--Chemisches
Institut, Universit\"at Heidelberg,
Im Neuenheimer Feld 229, D-69120 Heidelberg, Germany}

\author{Peter Schmelcher}
\affiliation{Theoretische Chemie, Physikalisch--Chemisches
Institut, Universit\"at Heidelberg,
Im Neuenheimer Feld 229, D-69120 Heidelberg, Germany}
\affiliation{Physikalisches Institut, Universit\"at Heidelberg,
Philosophenweg 12, D-69120 Heidelberg, Germany}

\date{\today}

\begin{abstract}
The formation of ultracold molecules via stimulated emission followed by
a radiative deexcitation cascade in the presence of a static electric
field is investigated. By analyzing the corresponding cross sections, we
demonstrate the possibility to populate the lowest rotational
excitations via photoassociation. The modification of the radiative
cascade due to the electric field leads to narrow rotational state
distributions in the vibrational ground state. External fields might
therefore represent an additional valuable tool towards the ultimate
goal of quantum state preparation of molecules. 
\end{abstract}


\maketitle

Over the past decade investigations of ultracold quantum gases have
been revealing a wealth of intriguing phenomena. In particular,
ultracold molecular systems represent a paradigm 
including molecular Bose-Einstein condensates \cite{jochim03,greiner03,zwierlein03}.
External fields are equally important for the preparation and control of ultracold systems.
They are used for cooling and trapping as well as for quantum state preparation or 
the tuning of atomic/molecular interactions, e.g., via Feshbach resonances \cite{wieman1999,pethick2002}.
Moreover, they provide a tool to manipulate chemical reactions and collisions \cite{krems05,krems06},
collisional spin relaxation in cold molecules \cite{krems06_2}, and
rovibrational spectra and transitions \cite{gonzalez06,mayle06}. 
The quest for ultracold molecules being prepared in well-defined rovibrational quantum
states is motivated by major perspectives such as the creation of quantum gases
with novel many-body properties, ultracold state-to-state chemical reaction dynamics,
or molecular quantum computation \cite{weidemueller2003,demille02,yelin06}.

A special focus is the study of heteronuclear polar dimers and the
possibility to create dipolar quantum gases \cite{speciss2004}. Due to the
permanent electric dipole moments of the constituent molecules, the
intermolecular interaction becomes long-ranged thereby
introducing exceptional properties for, e.g., the corresponding quantum
gases.
Electric fields play here an important role due to their immediate impact on
polar systems which leads to rotational orientation and alignment of 
the molecules.
The control of the formation process of ultracold polar molecules is a key
issue to arrive at molecular quantum gases. Photoassociation of ultracold
atoms is a widespread technique to produce ultracold molecules
\cite{jones:483}. Indeed, the formation via photoassociation of KRb
\cite{wang04}, LiCs \cite{kraft06}, NaCs \cite{haimberger04} and RbCs  
\cite{kerman04,sage:203001} has been reported recently.

Let us consider a mixture of two atomic species in their electronic  ground
states, exposed to a linearly polarized laser with a frequency corresponding to
the transition from the continuum to a highly excited vibrational bound 
state of the electronic ground state. 
The corresponding process of stimulated emission of a single photon allows for the
formation of a polar molecule, see fig.~1 
(note that hyperfine states are not considered).
It has been recently studied for the LiH and NaH molecules
\cite{cote06,juarros06} focusing on the ultracold regime where $s\to p$-wave
transitions dominate and only rovibrational bound states with angular
momentum equal to one can be produced. 
Subsequently, a radiative cascade of rovibrational transitions leads to the
vibrational ground state, thereby populating a broad range of rotational
molecular states. 

\begin{figure}
\includegraphics[width=8cm]{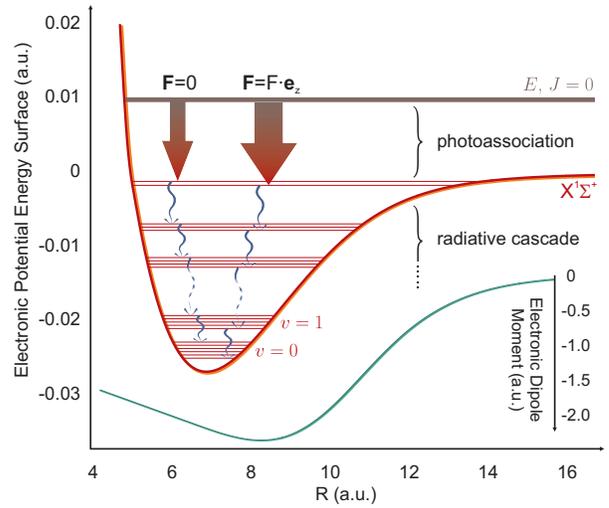}
\caption{\label{fig:pec}
(Color online)
Electronic potential energy curve and dipole moment function
of the electronic ground state $\textrm{X}\,^1\Sigma^+$ of the LiCs
molecule, in atomic units. 
The schematics of the OPA and RDC are also shown (not to scale).}
\end{figure}

In the present work we explore the impact of an additional homogeneous
static
electric  
field on the formation process of polar dimers. As indicated above, the latter
consists of the one-photon stimulated association (OPA) followed by a 
radiative deexcitation cascade (RDC). We demonstrate that the 
presence of the static field allows one to populate via the OPA
electrically dressed states evolving from  
field-free levels with zero angular momentum. This in combination with the
following RDC  yields a final rotational state distribution within the lowest
vibrational band which is significantly narrower compared to the field-free
case. Consequently, a static electric field can serve as a tool to prepare a
molecular gas with only a limited number of well-defined molecular quantum
states being populated. This might be a helpful step towards the goal of
reaching quantum gases of molecules in their rovibrational ground states.

Let us assume that perturbation theory suffices for the description of
the interaction with the laser field and for the interaction of the
electronic
motion with the static electric field. However,
a nonperturbative treatment is indispensable for the impact of the electric field
on the nuclear dynamics. The Hamiltonian for the rovibrational motion of the diatomic system in the 
Born-Oppenheimer approximation reads
\begin{equation}
\label{eq:rotvib_hamiltonian}
H= -\frac{\hbar^2}{2\mu R^2} \frac{\partial}{\partial R}
\left(R^2\frac{\partial}{\partial R }\right)
+\frac{\mathbf{J}^2(\theta,\phi)}{2\mu R^2} + \varepsilon(R)-FD(R)\cos\theta
\end{equation}
where the molecule fixed frame with the origin at the center of
mass of the nuclei has been employed. ($R,\theta,\phi$) are the
internuclear distance and the Euler angles, respectively. $\mu$,  
$\mathbf{J}(\theta,\phi)$, $\varepsilon(R)$, $F$, and $D(R)$ are the reduced
mass of 
the nuclei, the rotational angular momentum, the field-free electronic
potential energy curve, the electric field strength, and the electronic dipole
moment function, respectively. The electric field is parallel to the $z$-axis 
of the laboratory frame.  In field-free space, each bound state of the
molecule is characterized by its vibrational, rotational, and magnetic
quantum numbers $(v,J,M)$.  In the presence of the electric field only the
magnetic quantum number $M$ is conserved. However, for reasons of
addressability we will refer to the electrically dressed states by means of
the corresponding field-free quantum numbers. 

In the framework of the dipole approximation, the cross section of the stimulated
emission process from the continuum $\Psi_E(\mathbf{R})$ to the rovibrational 
state $\Psi_{\alpha}(\mathbf{R})$ is given by 
\begin{equation}
  \label{eq:cross_section}
  \sigma=\frac{\pi (E - E_{\alpha})}{\hbar c\epsilon_0}
|\langle\Psi_E|D(R)\cos\theta|\Psi_{\alpha}\rangle|^2
\end{equation}
where $\alpha$ labels the bound state and $E, E_{\alpha}$ 
are the continuum and rovibrational energies, respectively. 
The continuum wavefunctions are energy normalized while the rovibrational ones are ${\cal L}^2$ normalized.

The nuclear equation of motion associated with the Hamiltonian 
(\ref{eq:rotvib_hamiltonian}) is solved by means of a hybrid computational 
approach together with a Krylov-type diagonalization technique
\cite{arpackguide98}. 
For the vibrational coordinate we use a mapped discrete variable 
representation based on sine functions \cite{gonzalez06_2} in
a box of size $L$.
A  basis set expansion in terms of  the associated Legendre functions is 
performed for the angular coordinate.  This computational technique provides
an accurate description of highly  
excited bound states, while the continuum spectrum is 
discretized and described by ${\cal L}^2$ normalized states.
We emphasize that the traditional technique to represent energy normalized
continuum wave functions by means of ${\cal L}^2$ normalized functions
\cite{luckoenig04} is here not applicable due to the field-induced
coupling between the rotational and vibrational motions.
Moreover, the coupling significantly alters the definition of the density 
of continuum states. This issue has been solved by employing the
time-dependent formalism. Since our study focuses on the ultracold regime with
temperatures below $1$ mK, very large discretization boxes are required to
properly describe the continuum. 

Being a topical example, we focus on the polar alkali dimer LiCs.
The potential energy curve of its $\textrm{X}\,^1\Sigma^+$ electronic ground state is taken 
from experimental data \cite{staanum06} and the
electric dipole moment function from semi-empirical calculations
\cite{aymar05} (see fig.~1).
Continuum energies corresponding to $T=10, 100, 500\ \mu$K are considered.
We analyze the stimulated emission process for the population of the 
rovibrational levels $(44,J,0)$, which is a reasonable but robust
choice, with $J=0, 1$, requiring 
laser wavelengths of $181.13\,\mu$m and $181.56\,\mu$m, respectively.
Similar results are obtained for neighbouring vibrational bands.
The dependence of the total cross sections on the
electric field strength is shown in fig.~2. 
It is worth noting that the considered regime of field strengths covers
the experimentally accessible range and beyond: Static fields
$F>4\cdot 10^{-5}$ a.u. $= 200$  kV/cm are difficult to achieve in the laboratory. 
The cross sections increase with increasing 
temperature due to a larger overlap between the continuum and bound states.
In the absence of the electric field, the
ultracold regime is dominated by $s\to p$-wave transitions and the cross
section 
for populating the $(44,0,0)$ state via the OPA process is several orders of
magnitude 
smaller than the corresponding cross section involving the $(44,1,0)$ state.
\begin{figure}
\includegraphics[width=8cm]{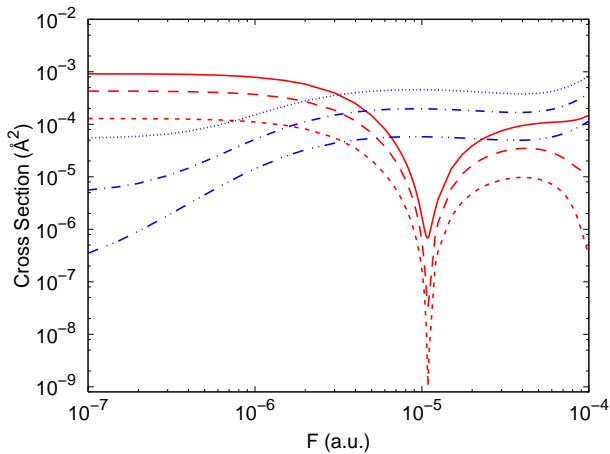}
\caption{\label{fig:sigma_versus_f}
(Color online)
Stimulated emission cross sections as a function of the electric field
strength 
for the states emerging from field-free levels with quantum numbers 
$(44,0,0)$ (dotted, dotted-dashed and double dotted-dashed lines) and
$(44,1,0)$ (solid, dashed and short dashed lines) for continuum energies  
corresponding to $T=500, \, 100$ and $10\,\mu$K, respectively.}
\end{figure}

Augmenting the field strength, the cross sections change significantly which
can be explained by the hybridization of the angular motion. For
$F\gtrsim 3\cdot 10^{-6}$ a.u., the cross section for the $(44,0,0)$ state 
becomes larger than the one belonging to the $(44,1,0)$ state.
Adopting $T=10\,\mu$K, the cross section of the $J=0$ state increases
monotonically by two orders of magnitude
within the regime $F = 10^{-7} - 10^{-5}$ a.u. 
Further increasing the field strength, it exhibits a plateau followed by a
weakly pronounced minimum and a strong increase thereafter.
In contrast to this, the cross section of the $(44,1,0)$ state shows a plateau
in the weak field regime.  
For stronger fields it rapidly decreases and reaches a deep minimum for 
$F=1.1 \cdot 10^{-5}$ a.u., 
followed by a broad maximum and a significant decrease for strong fields. 
This behaviour of the cross sections is due to the dominance of $s\to p$-wave
transitions in the ultracold regime combined with the hybridization of the
angular motion of the rovibrational state. Indeed, the presence of the
plateaus in both cross sections is accompanied by an analogous constancy 
of the contribution of the $p$-wave to the corresponding wave functions.
Similarly, the minimum for the $J=1$ state at $F=1.1 \cdot 10^{-5}$ a.u.\ is
due to the dominance of $s$, $d$, and $f$-waves along with a rather weak 
contribution of the $p$-wave. 

Let us shortly analyze the hybridization of the angular motion for these
two states in more detail. The (44,0,0) state has a high-field-seeking character with an
increasing expectation value $\langle\mathbf{J}^2\rangle$ for augmenting fields, i.e., the contribution
of higher field-free rotational states becomes larger as the electric
field strength is enhanced. Consequently, this level shows a pronounced 
angular momentum hybridization 
$\mathbf{J}^2_h(44,0,0)= \langle\mathbf{J}^2\rangle_{(44,0,0)}-J(J+1)=2.08$ 
and $7.88$, for $F=10^{-5}$ a.u.\ and $10^{-4}$ a.u., respectively, with 
$J$ being the corresponding field-free rotational quantum number.
On the other hand, the $(44,1,0)$ state shows initially a low-field-seeking 
behaviour where the admixing of lower rotations dominates and $\langle\mathbf{J}^2\rangle$ decreases:
$\mathbf{J}^2_h(44,1,0)= -0.11$ for $F=10^{-6}$ a.u. 
Further enhancing the electric field, this level becomes a high-field-seeker,
and $\langle\mathbf{J}^2\rangle$ increases after reaching a minimum. 
In the strong field regime, this level exhibits a very pronounced
hybridization $\mathbf{J}^2_h(44,1,0)=21.45$ for $F=10^{-4}$ a.u.

The formation rates in the OPA process can be varied
significantly by simply changing the intensity of the applied laser. 
Specifically, if we assume an atomic density $n = 10^{12}$
$\mathrm{cm}^{-3}$, a volume
$V= 10^{-6}$ $\mathrm{cm}^3$ illuminated by a laser beam with intensity $1$
$\frac{\mathrm{kW}}{\mathrm{cm}^2}$, and a temperature $T=1$ mK we obtain a
formation rate of $10^4$  molecules per second. 
Recently, the formation of LiCs molecules in a two-species 
magneto-optical trap by a two-photon process has been reported 
\cite{kraft06}. The molecules are formed in the electronic ground state, but
an analysis of the final vibrational state distribution is not provided. 
With  densities of $10^{10}$ $\mathrm{cm}^{-3}$ and  $5 \cdot 10^{9}$ $\mathrm{cm}^{-3}$ for the Li and Cs atoms, 
respectively, and assuming a temperature of $100\, \mu$K, a molecular production
rate between $1.4\pm 0.8$ and $140\pm 80$ molecules per second is estimated.
Using these densities for the Li and Cs atoms, our theoretical results for 
the molecular formation rate via the OPA process are in the same order of
magnitude. 
The reverse process of one photon absorption leading to the
dissociation of the molecules 
is suppressed by the RDC (see below), although this is much more pronounced for
light hydrides \cite{cote06} compared to alkali dimers. It might be further
eliminated by applying (chirped) laser pulses. Effects due to vibrational
quenching can be neglected for not too large molecular densities.

Since the photoassociation process results in vibrationally highly excited 
states, a cascade of spontaneous emission processes will follow.
The overall transition probability per unit time reads
$\Gamma_{\alpha}=\sum_{v'J'M'}^{E_{\alpha'}<E_{\alpha}}\Gamma_{\alpha,\alpha'}$
where the summation includes all open decay channels, i.e., final levels  
$\alpha'$ with $M'\in\lbrace M,M\pm1\rbrace$. The corresponding transition
rates are  
\begin{equation}
\Gamma_{\alpha,\alpha'}=
\frac{\omega_{\alpha,\alpha'}^3}{3\pi\epsilon_0\hbar c^3}|\langle
\Psi_{\alpha}|D(R)f_{M'}(\theta)|\Psi_{\alpha'}\rangle|^2\label{eq:sp_em_1}
\end{equation}
with 
$f_{M}(\theta)=\cos\theta$ and  $f_{M\pm1}(\theta)=\sin\theta$.
$\hbar\omega_{\alpha,\alpha'}$ is the energy difference between the
initial and final states. The total and single-channel radiative lifetimes 
are $\tau_{\alpha}=(\Gamma_{\alpha})^{-1}$ and 
$\tau_{\alpha,\alpha'}= (\Gamma_{\alpha,\alpha'})^{-1}$, respectively.
Our aim is to analyze the final rotational state distribution in the lowest
vibrational band resulting from the RDC. Starting from an initial state we
therefore add up all decay paths resulting in a specific final state. This
provides us the final $J,M$-state distributions.

\begin{figure}
\includegraphics[width=8cm]{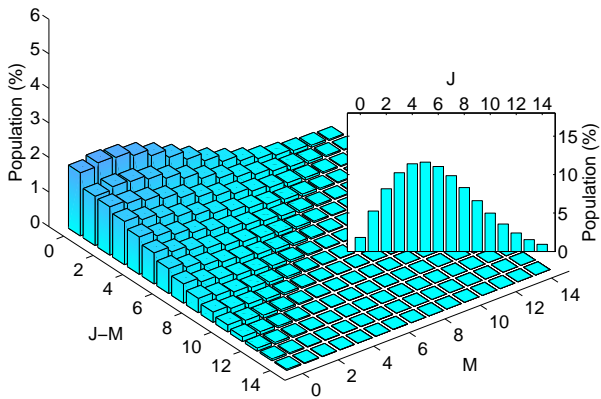}
\caption{\label{fig:distri_F_0}
(Color online)
Population distribution as a function of the magnetic quantum number $M$
and the difference $J-M$ for the decay cascade starting from the $(44,1,0)$
state in the absence of the electric field. The inset shows the cumulative 
population for fixed rotational quantum numbers.}
\end{figure}

Besides the possibility to form molecules evolving from field-free
states with
zero angular momentum 
by applying an electric field, the field interaction
significantly alters the radiative properties of the bound molecular states.
The rotational state distribution with varying $(J,M)$ in the ground
vibrational state which results from the radiation cascade of the $(44,1,0)$
state in the absence of the field is illustrated in fig.~3.
The inset shows the cumulative population for fixed rotational quantum numbers
$J$ ($|M|\le J$). 
Analogous results for the initial state $\alpha_0=(44,0,0)$, but for 
$F=4\cdot 10^{-5}$ a.u., are presented in fig.~4. 
It should be noted that negative magnetic quantum numbers obey the same
distribution  
due to the degeneracy of $\pm M$ states and the initial value $M=0$. 
Augmenting the field from zero to $F=4\cdot 10^{-5}$ a.u.,
corresponding to $200\frac{\mathrm{kV}}{\mathrm{cm}}$, 
the radiative lifetime of the $(44,0,0)$ state increases from $1.57$ s to
$2.15$ s, while for the $(44,1,0)$ state it is reduced from $1.87$ to
$1.76$ s.
In both cases, the prevailing variation of the vibrational quantum number for
each transition changes throughout the cascade from 
$\Delta v =6$ for the $v=44$ state to $\Delta v =1$ for low-lying vibrational 
states.
It is important to note that, in contrast to its favourable large
electric dipole moment, LiCs represents a molecule with quite long radiative
decay times. 
In comparison, heavier alkali dimers have even longer lifetimes, e.g.,
kiloseconds for the KRb molecule \cite{zemke04} while light hydrides such
as LiH exhibit much shorter lifetimes of the order of a few milliseconds
\cite{cote06}.

In the absence of the field, the selection rules $\Delta J=\pm 1$, 
$\Delta M\in\{0,\pm 1\}$ 
for dipole transitions hold, giving rise to a very broad
distribution of the population in the lowest vibrational band:
a large number of states exhibit a similar population. 
For a fixed $J$, the fully angular-momentum-polarized states always possess
the largest population, and for fixed $M$ the population slowly decreases with
increasing $J$. Moreover, the two fastest paths populate the rotationally
highly excited  $(0,20,\pm 20)$ states in $159$ s. The cumulative population
of the rotational bands exhibits a broad distribution with a maximum at
$J=5$. Similar results are obtained for the $(44,0,0)$ state. 
Once the vibrational ground state is populated, the rotational cascade is an
extremely slow process: The lifetimes of the $(0,1,\pm 1)$ and $(0,5,\pm 5)$ 
levels are $1.88\cdot 10^{6}$ s and $1.1\cdot 10^{4}$ s, respectively.
We remark that heating due to the random direction of the emitted photons
in the course of the RDC is a minor effect amounting to several tens of
nanokelvins for a typical setup \cite{cote06}.

The population of the final rotational states in the lowest vibrational
band following the OPA and RDC changes drastically in the
presence of the electric field. Due to the hybridization of the angular
motion,
only the selection rule $\Delta M\in\{0,\pm 1\}$ for the magnetic
quantum number
holds and many new transitions are possible. In particular, for fully polarized
angular momentum states the importance of purely vibrational transitions 
$\Delta J = \Delta M =0$
increases with increasing field strength and these transitions become dominant
in the strong field regime \cite{mayle06}. Consequently, the distribution 
is much narrower and exhibits pronounced peaks for a few fully
angular-momentum-polarized states as can be seen by comparing figs.~3 and 4.
For $F=4\cdot 10^{-5}$ a.u., the RDC is dominated by these purely vibrational 
transitions and the rovibrational ground state possesses the largest
population of $5.75\, \%$ compared to $2.02\, \%$ for $F=0$. This effect can
be further enhanced by employing even higher fields.
In addition, the rovibrational ground state is populated by the most
probable paths. Accordingly, we observe in the inset of fig.~4 that the
maximum of the cumulative  rotational state population is shifted to the $J=2$
band and includes $17.4\%$ of the overall norm. The entire RDC process to the
vibrational ground state typically takes several minutes. We remark that the
subsequent rotational decay itself is enhanced significantly in the presence
of the field and  the corresponding fully angular-momentum-polarized states
possess a  unique radiative decay route to the rovibrational ground state
\cite{mayle06}. 
\begin{figure}
\includegraphics[width=8cm]{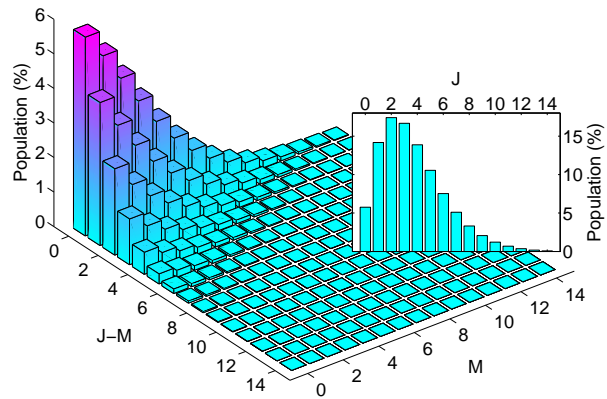}
\caption{\label{fig:distri_F_4e-5}
(Color online)
Same as in fig.~\ref{fig:distri_F_0} but with $(44,0,0)$ as initial
state 
and for $F=4\cdot 10^{-5}$ a.u.}
\end{figure}

The above analysis demonstrates that the formation of ultracold molecules in
their electronic ground state via a single-photon stimulated emission
process followed by a radiative cascade can be significantly altered by the
presence of a static electric field. The resulting narrow rotational
distributions in the lowest vibrational band represent a remarkable step
towards the ultimate goal of degenerate quantum gases of molecules in their
rovibrational ground state, or more generally in  any desired unique quantum
state. Although our study focuses on the LiCs molecule, a qualitatively similar
behaviour has to be expected for other heteronuclear dimers depending on the
corresponding potential energy curve and electric dipole moment function. 

R.G.F.\ acknowledges the support of the Junta de Andaluc\'{\i}a under
the program of Retorno de Investigadores a Centros de Investigaci\'on
Andaluces.
Financial support by the Acciones Integradas Hispano Alemanas 
HA2005--0038 (MEC and DAAD) and of the Spanish projects FIS2005--00973 (MEC) as
well as FQM--0207 and FQM--481 (Junta de Andaluc\'{\i}a) is gratefully
appreciated. We thank particularly Hans-Dieter Meyer for his strong support
concerning the workout of the time-dependent formalism. Roland Wester is
acknowledged for fruitful discussions, Peter Staanum for providing us with the
potential energy curve, and Olivier Dulieu for the electric dipole moment of
the LiCs molecule.


\begin{thebibliography}{28}
\expandafter\ifx\csname natexlab\endcsname\relax\def\natexlab#1{#1}\fi
\expandafter\ifx\csname bibnamefont\endcsname\relax
  \def\bibnamefont#1{#1}\fi
\expandafter\ifx\csname bibfnamefont\endcsname\relax
  \def\bibfnamefont#1{#1}\fi
\expandafter\ifx\csname citenamefont\endcsname\relax
  \def\citenamefont#1{#1}\fi
\expandafter\ifx\csname url\endcsname\relax
  \def\url#1{\texttt{#1}}\fi
\expandafter\ifx\csname urlprefix\endcsname\relax\def\urlprefix{URL }\fi
\providecommand{\bibinfo}[2]{#2}
\providecommand{\eprint}[2][]{\url{#2}}

\bibitem[{\citenamefont{{S. Jochim \emph{et al.}}}(2003)}]{jochim03}
\bibinfo{author}{\bibnamefont{{S. Jochim \emph{et al.}}}},
  \bibinfo{journal}{Science} \textbf{\bibinfo{volume}{302}},
  \bibinfo{pages}{2101} (\bibinfo{year}{2003}).

\bibitem[{\citenamefont{Greiner et~al.}(2003)\citenamefont{Greiner,
Regal, and
  Jin}}]{greiner03}
\bibinfo{author}{\bibfnamefont{M.}~\bibnamefont{Greiner}},
  \bibinfo{author}{\bibfnamefont{C.~A.} \bibnamefont{Regal}},
\bibnamefont{and}
  \bibinfo{author}{\bibfnamefont{D.~S.} \bibnamefont{Jin}},
  \bibinfo{journal}{Nature} \textbf{\bibinfo{volume}{426}},
  \bibinfo{pages}{537} (\bibinfo{year}{2003}).

\bibitem[{\citenamefont{{M. W. Zwierlein \emph{et
al.}}}(2003)}]{zwierlein03}
\bibinfo{author}{\bibnamefont{{M. W. Zwierlein \emph{et al.}}}},
  \bibinfo{journal}{Phys. Rev. Lett.} \textbf{\bibinfo{volume}{91}},
  \bibinfo{pages}{250401} (\bibinfo{year}{2003}).

\bibitem[{\citenamefont{Wieman et~al.}(1999)\citenamefont{Wieman,
Pritchard,
  and Wineland}}]{wieman1999}
\bibinfo{author}{\bibfnamefont{C.}~\bibnamefont{Wieman}},
  \bibinfo{author}{\bibfnamefont{D.~E.} \bibnamefont{Pritchard}},
  \bibnamefont{and}
\bibinfo{author}{\bibfnamefont{D.}~\bibnamefont{Wineland}},
  \bibinfo{journal}{Rev. Mod. Phys.} \textbf{\bibinfo{volume}{71}},
  \bibinfo{pages}{S253} (\bibinfo{year}{1999}).

\bibitem[{\citenamefont{Pethick and Smith}(2002)}]{pethick2002}
\bibinfo{author}{\bibfnamefont{C.}~\bibnamefont{Pethick}}
\bibnamefont{and}
  \bibinfo{author}{\bibfnamefont{H.}~\bibnamefont{Smith}},
  \bibinfo{journal}{Bose-Einstein Condensation in Dilute Gases,
Cambridge
  University Press}  (\bibinfo{year}{2002}).

\bibitem[{\citenamefont{Krems}(2005)}]{krems05}
\bibinfo{author}{\bibfnamefont{R.~V.} \bibnamefont{Krems}},
  \bibinfo{journal}{Int. Rev. Phys. Chem.}
\textbf{\bibinfo{volume}{24}},
  \bibinfo{pages}{99} (\bibinfo{year}{2005}).

\bibitem[{\citenamefont{Krems}(2006)}]{krems06}
\bibinfo{author}{\bibfnamefont{R.~V.} \bibnamefont{Krems}},
  \bibinfo{journal}{Phys. Rev. Lett.} \textbf{\bibinfo{volume}{96}},
  \bibinfo{pages}{123202} (\bibinfo{year}{2006}).

\bibitem[{\citenamefont{Tscherbul and Krems}(2006)}]{krems06_2}
\bibinfo{author}{\bibfnamefont{T.~V.} \bibnamefont{Tscherbul}}
  \bibnamefont{and} \bibinfo{author}{\bibfnamefont{R.~V.}
\bibnamefont{Krems}},
  \bibinfo{journal}{Phys. Rev. Lett.} \textbf{\bibinfo{volume}{97}},
  \bibinfo{pages}{083201} (\bibinfo{year}{2006}).

\bibitem[{\citenamefont{Gonz\'alez-F\'erez
  et~al.}(2006)\citenamefont{Gonz\'alez-F\'erez, Mayle, and
  Schmelcher}}]{gonzalez06}
\bibinfo{author}{\bibfnamefont{R.}~\bibnamefont{Gonz\'alez-F\'erez}},
  \bibinfo{author}{\bibfnamefont{M.}~\bibnamefont{Mayle}},
\bibnamefont{and}
  \bibinfo{author}{\bibfnamefont{P.}~\bibnamefont{Schmelcher}},
  \bibinfo{journal}{Chem. Phys.} \textbf{\bibinfo{volume}{329}},
  \bibinfo{pages}{203} (\bibinfo{year}{2006}).

\bibitem[{\citenamefont{Mayle et~al.}(2007)\citenamefont{Mayle,
  Gonz\'alez-F\'erez, and Schmelcher}}]{mayle06}
\bibinfo{author}{\bibfnamefont{M.}~\bibnamefont{Mayle}},
  \bibinfo{author}{\bibfnamefont{R.}~\bibnamefont{Gonz\'alez-F\'erez}},
  \bibnamefont{and}
  \bibinfo{author}{\bibfnamefont{P.}~\bibnamefont{Schmelcher}},
  \bibinfo{journal}{Phys. Rev. A} \textbf{\bibinfo{volume}{75}},
  \bibinfo{pages}{013421} (\bibinfo{year}{2007}).

\bibitem[{\citenamefont{DeMille}(2002)}]{demille02}
\bibinfo{author}{\bibfnamefont{D.}~\bibnamefont{DeMille}},
  \bibinfo{journal}{Phys. Rev. Lett.} \textbf{\bibinfo{volume}{88}},
  \bibinfo{pages}{067901} (\bibinfo{year}{2002}).

\bibitem[{\citenamefont{Yelin et~al.}(2006)\citenamefont{Yelin, Kirby,
and
  C\^ot\'e}}]{yelin06}
\bibinfo{author}{\bibfnamefont{S.~F.} \bibnamefont{Yelin}},
  \bibinfo{author}{\bibfnamefont{K.}~\bibnamefont{Kirby}},
\bibnamefont{and}
  \bibinfo{author}{\bibfnamefont{R.}~\bibnamefont{C\^ot\'e}},
  \bibinfo{journal}{Phys. Rev. A} \textbf{\bibinfo{volume}{74}},
  \bibinfo{pages}{050301(R)} (\bibinfo{year}{2006}).

\bibitem[{\citenamefont{Weidem\"uller and {C. Zimmermann,
  Eds.}}(2003)}]{weidemueller2003}
\bibinfo{author}{\bibfnamefont{M.}~\bibnamefont{Weidem\"uller}}
  \bibnamefont{and} \bibinfo{author}{\bibnamefont{{C. Zimmermann,
Eds.}}},
  \bibinfo{journal}{Interactions in Ultracold Gases, Wiley-VCH}
  (\bibinfo{year}{2003}).

\bibitem[{spe(2004)}]{speciss2004}
\emph{\bibinfo{title}{{Topical Issue on Ultracold Polar Molecules:
Formation
  and Collisions}}}, \bibinfo{howpublished}{Eur. Phys. J. D 31}
  (\bibinfo{year}{2004}).

\bibitem[{\citenamefont{Jones et~al.}(2006)\citenamefont{Jones,
Tiesinga, Lett,
  and Julienne}}]{jones:483}
\bibinfo{author}{\bibfnamefont{K.~M.} \bibnamefont{Jones}},
  \bibinfo{author}{\bibfnamefont{E.}~\bibnamefont{Tiesinga}},
  \bibinfo{author}{\bibfnamefont{P.~D.} \bibnamefont{Lett}},
\bibnamefont{and}
  \bibinfo{author}{\bibfnamefont{P.~S.} \bibnamefont{Julienne}},
  \bibinfo{journal}{Rev. Mod. Phys.} \textbf{\bibinfo{volume}{78}},
  \bibinfo{pages}{483} (\bibinfo{year}{2006}).

\bibitem[{\citenamefont{{D. Wang \emph{et al.}}}(2004)}]{wang04}
\bibinfo{author}{\bibnamefont{{D. Wang \emph{et al.}}}},
\bibinfo{journal}{Eur.
  Phys. J. D} \textbf{\bibinfo{volume}{31}}, \bibinfo{pages}{165}
  (\bibinfo{year}{2004}).

\bibitem[{\citenamefont{{S. D. Kraft \emph{et al.}}}(2006)}]{kraft06}
\bibinfo{author}{\bibnamefont{{S. D. Kraft \emph{et al.}}}},
  \bibinfo{journal}{J. Phys. B} \textbf{\bibinfo{volume}{39}},
  \bibinfo{pages}{S993} (\bibinfo{year}{2006}).

\bibitem[{\citenamefont{Haimberger
et~al.}(2004)\citenamefont{Haimberger,
  Kleinert, Bhattacharya, and Bigelow}}]{haimberger04}
\bibinfo{author}{\bibfnamefont{C.}~\bibnamefont{Haimberger}},
  \bibinfo{author}{\bibfnamefont{J.}~\bibnamefont{Kleinert}},
  \bibinfo{author}{\bibfnamefont{M.}~\bibnamefont{Bhattacharya}},
  \bibnamefont{and} \bibinfo{author}{\bibfnamefont{N.~P.}
  \bibnamefont{Bigelow}}, \bibinfo{journal}{Phys. Rev. A}
  \textbf{\bibinfo{volume}{70}}, \bibinfo{pages}{021402(R)}
  (\bibinfo{year}{2004}).

\bibitem[{\citenamefont{Kerman et~al.}(2004)\citenamefont{Kerman, Sage,
Sainis,
  Bergeman, and DeMille}}]{kerman04}
\bibinfo{author}{\bibfnamefont{A.~J.} \bibnamefont{Kerman}},
  \bibinfo{author}{\bibfnamefont{J.~M.} \bibnamefont{Sage}},
  \bibinfo{author}{\bibfnamefont{S.}~\bibnamefont{Sainis}},
  \bibinfo{author}{\bibfnamefont{T.}~\bibnamefont{Bergeman}},
\bibnamefont{and}
  \bibinfo{author}{\bibfnamefont{D.}~\bibnamefont{DeMille}},
  \bibinfo{journal}{Phys. Rev. Lett.} \textbf{\bibinfo{volume}{92}},
  \bibinfo{pages}{153001} (\bibinfo{year}{2004}).

\bibitem[{\citenamefont{Sage et~al.}(2005)\citenamefont{Sage, Sainis,
Bergeman,
  and DeMille}}]{sage:203001}
\bibinfo{author}{\bibfnamefont{J.~M.} \bibnamefont{Sage}},
  \bibinfo{author}{\bibfnamefont{S.}~\bibnamefont{Sainis}},
  \bibinfo{author}{\bibfnamefont{T.}~\bibnamefont{Bergeman}},
\bibnamefont{and}
  \bibinfo{author}{\bibfnamefont{D.}~\bibnamefont{DeMille}},
  \bibinfo{journal}{Phys. Rev. Lett.} \textbf{\bibinfo{volume}{94}},
  \bibinfo{pages}{203001} (\bibinfo{year}{2005}).

\bibitem[{\citenamefont{Juarros
  et~al.}(2006{\natexlab{a}})\citenamefont{Juarros, Pellegrini, Kirby,
and
  C\^ot\'e}}]{cote06}
\bibinfo{author}{\bibfnamefont{E.}~\bibnamefont{Juarros}},
  \bibinfo{author}{\bibfnamefont{P.}~\bibnamefont{Pellegrini}},
  \bibinfo{author}{\bibfnamefont{K.}~\bibnamefont{Kirby}},
\bibnamefont{and}
  \bibinfo{author}{\bibfnamefont{R.}~\bibnamefont{C\^ot\'e}},
  \bibinfo{journal}{Phys. Rev. A} \textbf{\bibinfo{volume}{73}},
  \bibinfo{pages}{041403(R)} (\bibinfo{year}{2006}{\natexlab{a}}).

\bibitem[{\citenamefont{Juarros
  et~al.}(2006{\natexlab{b}})\citenamefont{Juarros, Kirby, and
  C\^ot\'e}}]{juarros06}
\bibinfo{author}{\bibfnamefont{E.}~\bibnamefont{Juarros}},
  \bibinfo{author}{\bibfnamefont{K.}~\bibnamefont{Kirby}},
\bibnamefont{and}
  \bibinfo{author}{\bibfnamefont{R.}~\bibnamefont{C\^ot\'e}},
  \bibinfo{journal}{J. Phys. B} \textbf{\bibinfo{volume}{39}},
  \bibinfo{pages}{S965} (\bibinfo{year}{2006}{\natexlab{b}}).

\bibitem[{\citenamefont{Lehoucq et~al.}(1998)\citenamefont{Lehoucq,
Sorensen,
  and Yang}}]{arpackguide98}
\bibinfo{author}{\bibfnamefont{R.~B.} \bibnamefont{Lehoucq}},
  \bibinfo{author}{\bibfnamefont{D.~C.} \bibnamefont{Sorensen}},
  \bibnamefont{and}
\bibinfo{author}{\bibfnamefont{C.}~\bibnamefont{Yang}},
  \emph{\bibinfo{title}{ARPACK User's Guide}}
(\bibinfo{publisher}{SIAM},
  \bibinfo{address}{Philadelphia}, \bibinfo{year}{1998}).

\bibitem[{\citenamefont{Gonz\'alez-F\'erez and Meyer}()}]{gonzalez06_2}
\bibinfo{author}{\bibfnamefont{R.}~\bibnamefont{Gonz\'alez-F\'erez}}
  \bibnamefont{and} \bibinfo{author}{\bibfnamefont{H.-D.}
\bibnamefont{Meyer}},
  \bibinfo{note}{{preprint 2006}}.

\bibitem[{\citenamefont{Luc-Koenig
et~al.}(2004)\citenamefont{Luc-Koenig,
  Vatasescu, and Masnou-Seeuws}}]{luckoenig04}
\bibinfo{author}{\bibfnamefont{E.}~\bibnamefont{Luc-Koenig}},
  \bibinfo{author}{\bibfnamefont{M.}~\bibnamefont{Vatasescu}},
  \bibnamefont{and}
  \bibinfo{author}{\bibfnamefont{F.}~\bibnamefont{Masnou-Seeuws}},
  \bibinfo{journal}{Eur. Phys. J. D} \textbf{\bibinfo{volume}{31}},
  \bibinfo{pages}{239} (\bibinfo{year}{2004}).

\bibitem[{\citenamefont{Staanum et~al.}(2006)\citenamefont{Staanum,
Pashov,
  Knoeckel, and Tiemann}}]{staanum06}
\bibinfo{author}{\bibfnamefont{P.}~\bibnamefont{Staanum}},
  \bibinfo{author}{\bibfnamefont{A.}~\bibnamefont{Pashov}},
  \bibinfo{author}{\bibfnamefont{H.}~\bibnamefont{Knoeckel}},
\bibnamefont{and}
  \bibinfo{author}{\bibfnamefont{E.}~\bibnamefont{Tiemann}},
  \bibinfo{journal}{physics/0612031}  (\bibinfo{year}{2006}).

\bibitem[{\citenamefont{Aymar and Dulieu}(2005)}]{aymar05}
\bibinfo{author}{\bibfnamefont{M.}~\bibnamefont{Aymar}}
\bibnamefont{and}
  \bibinfo{author}{\bibfnamefont{O.}~\bibnamefont{Dulieu}},
  \bibinfo{journal}{J. Chem. Phys.} \textbf{\bibinfo{volume}{122}},
  \bibinfo{pages}{204302} (\bibinfo{year}{2005}).

\bibitem[{\citenamefont{Zemke and Stwalley}(2004)}]{zemke04}
\bibinfo{author}{\bibfnamefont{W.~T.} \bibnamefont{Zemke}}
\bibnamefont{and}
  \bibinfo{author}{\bibfnamefont{W.~C.} \bibnamefont{Stwalley}},
  \bibinfo{journal}{J. Chem. Phys.} \textbf{\bibinfo{volume}{120}},
  \bibinfo{pages}{88} (\bibinfo{year}{2004}).

\end{thebibliography}
\end{document}